\documentclass[aps,prb,twocolumn,superscriptaddress,floatfix]{revtex4}
\usepackage{graphicx}
\usepackage{epsfig}
\usepackage{amsmath}
\usepackage{amssymb}
\DeclareGraphicsExtensions{.jpg, .jpg, .eps, .pdf}
\graphicspath{{figures/}}

\begin{document}

\title{Static disorder and local structure in zinc(II) isonicotinate,\\ a quartzlike metal--organic framework}

\author{Ines E. Collings}
\affiliation{Department of Chemistry, University of Oxford, Inorganic Chemistry Laboratory, South Parks Road, Oxford OX1 3QR, U.K.}

\author{Matthew G. Tucker}
\affiliation{ISIS Facility, Rutherford Appleton Laboratory, Harwell Science and Innovation Campus, Didcot OX11 0QX, U.K.}

\author{David A. Keen}
\affiliation{ISIS Facility, Rutherford Appleton Laboratory, Harwell Science and Innovation Campus, Didcot OX11 0QX, U.K.}

\author{Andrew L. Goodwin}
\email[]{andrew.goodwin@chem.ox.ac.uk}
\affiliation{Department of Chemistry, University of Oxford, Inorganic Chemistry Laboratory, South Parks Road, Oxford OX1 3QR, U.K.}

\date{\today}
\begin{abstract}
Using a combination of Rietveld and RMC refinement of neutron total scattering data, we find that the 10\,K structure of zinc(II) isonicotinate shows strong evidence of static disorder. This disorder takes the form of transverse displacements of the isonicotinate ligand and results in elongated atomic displacement parameters and dampened oscillations of the experimental $G(r)$. We analyse the RMC configurations using an approach derived from geometric algebra. Complications regarding the inclusion of hydrogenous guest molecules within the pore structure are discussed. This study highlights the way in which structural flexibility can give rise to multiple low-energy ground states in MOF-type materials.
\end{abstract}

\maketitle

\section{Introduction}

The structural analogies between traditional oxide-containing frameworks and the class of materials known as metal--organic frameworks (MOFs) are certainly well established.\cite{Hoskins_1990} A topical example is the family of zeolitic imidazolate frameworks (ZIFs), which are MOF analogues of zeolitic SiO$_2$ polymorphs.\cite{Lehnert_1980,Tian_2003,Banerjee_2008} Replacing Si$^{4+}$ and O$^{2-}$ ions by tetrahedral Zn$^{2+}$ centres and bridging imidazolate linkers respectively, it is possible to generate a wide range of zeolite-like framework structures with increased pore volume. Like zeolites, ZIFs offer attractive gas storage and catalytic functionality,\cite{Banerjee_2008} but they couple these properties with the characteristic chemical versatility for which MOFs are renowned (\emph{e.g.}\ the imidazolate linker can be replaced by substituted derivatives and the transition metal dication varied). The extensive range of functionalities exhibited by ZIFs and many other MOF families has generated a strong and sustained interest in studying structure/property relationships in such `hybrid' inorganic-organic materials;\cite{Cheetham:2007kx,Ma:2010vn,Dybtsev:2006ys,Halder:2002zr,Qiu:2009ly} consequently, there is a great deal now known regarding their structural chemistry.

What has become increasingly evident is that the structural behaviour of MOFs is often significantly more complex than might have been anticipated. A surprisingly large number---including the canonical MOF-5 (Ref.~\onlinecite{Li_1999})---exhibit negative thermal expansion (NTE; \emph{i.e.}, their lattices expand on cooling).\cite{Zhou:2008uq, Wu:2008kx} Some undergo pressure- and temperature-induced amorphisation processes.\cite{Bennett:2010fk} Yet others show unprecedented structural flexibility, being capable of dramatic changes in crystallite dimensions during absorption or desorption of guest species.\cite{Serre:2002vn} While it is the case that similar mechanical phenomena have been documented for oxide-containing frameworks (\emph{e.g.}\ NTE in ZrW$_2$O$_8$, see Ref.~\onlinecite{Mary:1996}), the effects observed in MOFs are almost always much more extreme. The increased structural flexibility of molecular metal--ligand--metal linkages is heavily implicated in the fundamental difference in magnitude of behaviour:  metal--ligand geometries are more easily distorted when the ligand is large and flexible than when it is a single oxide ion.

The prevalence of low-energy deformation mechanisms in MOFs should result in an increased propensity for structural distortion, which if incoherent could resemble the static disorder often found in zeolites\cite{Wragg:2008} and perovskite frameworks\cite{Rodriguez:2009}---or when coherent may give rise to an extreme symmetry lowering observed for some complex framework oxides.\cite{Lister:2004} Yet it seems that reports of static disorder in MOFs are relatively few in number: there is, for example, some discussion of ligand orientation disorder and strongly anisotropic displacement parameters in Cu(4-oxopyrimidinate)$_2$;\cite{Barea_2003} likewise molecular dynamics studies of MOF-5 and its isoreticular congeners have also pointed to the existence of low-barrier enthalpy landscapes thought to be responsible for both static and dynamic disorder.\cite{Amirjalayer:2008fk,Jhon:2007uq} A traditional emphasis within the field on average structure studies---which are also usually performed at temperatures insufficiently low to distinguish static disorder from vibrational motion---has perhaps meant that the degree of static disorder within MOFs has remained poorly understood.

Here, as part of a broader study into the local structure and dynamics of mineralomimetic MOFs, we report a combined average- and local-structure investigation into the existence and nature of static disorder in zinc(II) isonicotinate, Zn(ISN)$_2$. Our approach has been to collect neutron total scattering data for Zn(ISN)$_2$ at 10\,K, which we analyse using Rietveld refinement (average structure) and reverse Monte Carlo (RMC) refinement of the corresponding pair distribution function (PDF) data (local structure). We find strong evidence in both types of refinement for the existence of static disorder that resembles local transverse displacements of the isonicotinate ligands.

\begin{figure}
\includegraphics{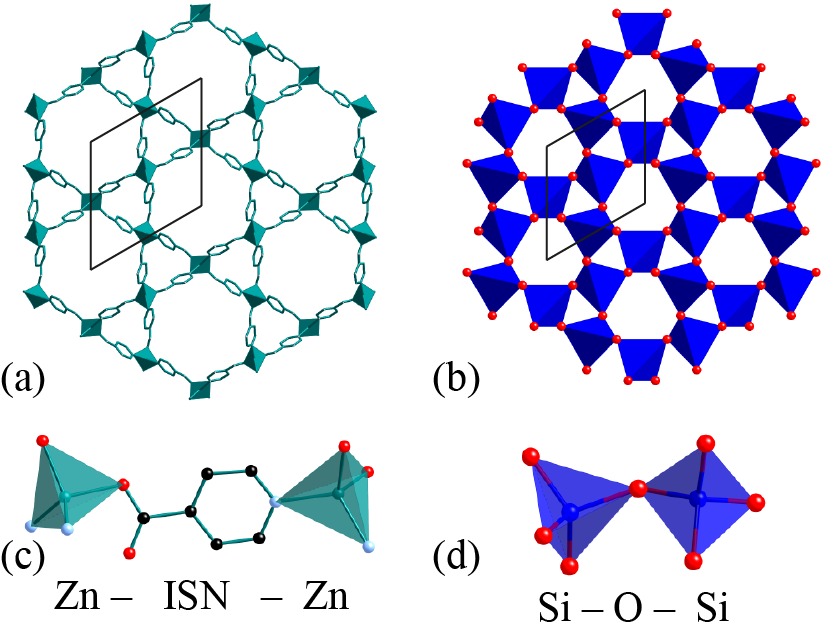}
\caption{Crystal structure of (a) Zn(ISN)$_2$ with [ZnN$_2$O$_2$] tetrahedra connected via isonicotinate ligands, and (b) $\alpha$-quartz with connected [SiO$_4$] tetrahedra.  Both structures are viewed down the \emph{c} axis with their unit cell shown.  Local coordination environment of (c) Zn(ISN)$_2$ and (d) $\alpha$-quartz.  (Zn, green; N, light blue; O, red; C, black; Si, dark blue, and isonicotinate hydrogens have been removed for clarity).\label{fig1}}
\end{figure}

The Zn(ISN)$_2$ framework structure contains Zn$^{2+}$ centres that are fourfold-coordinated by isonicotinate anions, with each anion bridging two Zn$^{2+}$ cations to form a tetrahedral net with the quartz topology [Fig.~\ref{fig1}]. The coordination sphere of each Zn$^{2+}$ centre consists of two pyridinyl N donor atoms and two carboxylate O atoms---an arrangement that is inconsistent with the corresponding site symmetry imposed by the $\beta$-quartz structure. Instead, the material adopts the lower-symmetry $\alpha$-quartz structure, crystallising in the hexagonal space group $P6_2$.\cite{Wu:2009} As in quartz itself, the structure of Zn(ISN)$_2$ contains one-dimensional pores parallel to the crystallographic $\mathbf c$ axis. In as-prepared samples these pores are occupied by solvent molecules, but the pores can be evacuated by heating to 100\,$^\circ$C.\cite{Sun:2002} There remains some controversy as to whether the correct space group for Zn(ISN)$_2$ is actually $P3_1$, which is a subgroup of $P6_2$ and which demands the existence of two crystallographically distinct ISN units in the asymmetric unit.\cite{Wu:2009, Sun:2002} 

Our paper begins by describing the experimental methods used in our study, together with our approach to RMC refinements of the neutron PDF data we have collected. We note that this is somewhat of a test-case for the use of our {\sc rmcprofile} code\cite{Tucker:2007} in the study of structural disorder in crystalline MOFs, and we have encountered some difficulties through an unquantifiable degree of absorption of atmospheric H$_2$O into our deuterated sample. Mindful of the ensuing limitations, we present the results of our average-structure and RMC studies and discuss the influence of the included solvent on the conclusions drawn. We finish with a discussion on the nature of static disorder in this material in the context of correlated (thermal) displacements within $\alpha$-quartz itself.

\section{Materials and Methods}
\subsection{Synthesis}

A polycrystalline sample of zinc(II) isonicotinate was prepared by mixing stoichiometric quantities of zinc(II) acetate and $d^4$-isonicotinic acid (QMx, 98\% D) dissolved in dimethylsulfoxide (DMSO). The white powder formed on mixing was filtered, washed and dried \emph{in vacuo} (100\,$^{\circ}$C, 24\,h) in order to remove all solvent from the framework pores. Working within a glovebox, the dried sample (\emph{ca} 2\,g) was transferred to a vanadium can suitable for neutron total scattering measurements.

\subsection{Neutron total scattering}

Neutron total scattering data were collected at 10\,K for Zn(ISN)$_2$ using the time-of-flight diffractometer GEM at ISIS. \cite{Williams:1997,Day:2004,AlexC:2005} For the experiment, approximately 2\,g of Zn(ISN)$_2$, prepared as described above, was placed within a cylindrical vanadium can of 8.3\,mm diameter and 6\,cm height. This can was loaded at room temperature inside a closed cycle helium refrigerator and the temperature was lowered slowly to 10\,K. Total scattering data were collected over a large range of scattering vectors of magnitudes $0.6\leq Q\leq 40$\,\AA$^{-1}$, giving a real-space resolution of order $\Delta r\simeq 0.09$\,\AA.  

The total scattering data were corrected using standard methods, taking into account the effects of background scattering, absorption, multiple scattering within the sample, beam intensity variations, the Placzek inelastic correction, and hydrogen content corrections.\cite{Keen:2001fk}  These corrected data were then converted to experimental $G(r)$ and $F(Q)$ functions:\cite{Dove:2002kx,Keen:2001fk}
\begin{eqnarray}
F(Q)&=&\rho_0 \int_0^\infty4\pi r^2 G(r)\frac{\sin Qr}{Qr}\mathrm{d}r\label{fq}\\
G(r)&=&\sum_{i,j=1}^nc_ic_j\bar{b}_i\bar{b}_j[g_{ij}(r)-1],
\end{eqnarray}
where
\begin{equation}
g_{ij}(r)=\frac{n_{ij}(r)}{4\pi r^2\mathrm{d}r\rho_0},\label{gr}
\end{equation}
$n_{ij}(r)$ is the number of pairs of atoms of type $i$ and $j$ separated by distance $r$, $\rho_0$ is the number density, $c_i$ the concentration of each species $i$ and $b_i$ the corresponding neutron scattering length. The Bragg profiles for each data set were extracted from the scattering data collected using the detector banks centred on on scattering angles $2\theta = 34.96^{\circ}, 63.62^{\circ}, 91.30^{\circ},$ and $154.40^{\circ}$. 

Because the number density $\rho_0$ enters Eqs.~\eqref{fq} and \eqref{gr}, it is possible during data normalisation to check for consistency between the expected value of $\rho_0$ and the value for which normalisation is most robust. In the present case, a value of $\rho_0$ corresponding to guest-free Zn(ISN)$_2$ was found not to be the value most consistent with the observed data; moreover, by comparing the incoherent scattering levels associated with each detector bank it was possible to deduce that the sample as measured was actually contaminated with a significant quantity of hydrogenous material, despite the care taken to evacuate the framework completely. 

This unquantifiable degree of solvation had three consequences for our (usually quantitative) RMC modelling of the PDF data. First, we found that quantitative normalisation of the $G(r)$ function was not possible, and so for RMC refinement a smoothly-varying function was subtracted from the normalised data---this did not affect the final model but improved visually the quality of our fits to data. Second, it was not possible to fit the very lowest-$r$ region of the PDF since this was likely to contain a well-structured contribution from the included solvent (\emph{e.g.}\ the O--H distance if the solvent were H$_2$O). Third, because the value of $\rho_0$ was not known accurately it was necessary to fit the PDF only while allowing refinement of an overall scale parameter. We note that such an approach is quite common for other PDF fitting procedures.\cite{Farrow:2007}
  
\subsection{Average structure refinement}

The experimental Bragg diffraction profiles were fitted with the {\sc gsas} Rietveld refinement program\cite{VonDreele:2000,Toby:2001} using both the published $P3_1$ and $P6_2$ structural models.\cite{Sun:2002,Wu:2009} Atomic coordinates of the organic ligand were refined using rigid body constraints in order to minimise the number of refinable parameters; likewise a single set of anisotropic displacement parameters was refined for all atoms in the isonicotinate group. Zn atom coordinates were refined while fixing the $z$ component (there is no unique origin in $z$ for either $P3_1$ or $P6_2$) and an isotropic displacement parameter was allowed to refine freely. Solvent occupancy within the pore network was treated using a number of different approaches, and these are discussed in more detail in the results section below.

\subsection{Reverse Monte Carlo refinement}

RMC refinements were carried out using the {\sc rmcprofile} code.\cite{Tucker:2007} To the best of our knowledge, this study represents the first RMC study of a crystalline MOF and we detail below some of the difficulties we have encountered in the process. As for all RMC studies, the basic refinement objective is to produce atomistic configurations that fit simultaneously the experimental $G(r)$, $F(Q)$, and Bragg profile $I(t)$ functions.  This is achieved by accepting or rejecting random atomic moves produced by the Metropolis Monte Carlo algorithm, where in this case the Monte Carlo `energy' function is determined by the quality of the fits to data.  The motivation behind fitting both real-space $G(r)$ and reciprocal-space $F(Q)$/$I(t)$ functions is to probe local distortions in the framework in a manner that is inherently consistent with the long-range periodic order reflected in the Bragg intensities. Similar refinements of crystalline materials in which reciprocal-space data are not used have a tendency to become unphysically disordered,\cite{Tucker:2001} and we were keen to assess the degree of structural disorder in Zn(ISN)$_2$ on the most realistic level possible.

Our starting configurations for the RMC process were based on a orthogonal supercell related to the crystallographic cell by the transformation
\begin{equation}
\left[ \begin{matrix}
a \\ b \\ c 
\end{matrix}  \right]_{\textrm{RMC}} =
\left[ \begin{array}{ccc}
4 & 0 & 0  \\
2 & 4 & 0  \\
0 & 0 & 10  \\
\end{array}  \right] \times
\left[ \begin{matrix}
a \\ b \\ c  \\
\end{matrix} \right]_{P6_2},
\end{equation}
giving a cell of dimensions 60.14 \AA\ $\times$ 53.82 \AA\ $\times$ 61.36\,\AA.  The use in RMC of orthogonal axes for hexagonal systems means the configurations can be prepared with dimensions approximately equal in each direction. This maximises the pair distribution cut-off value $r_{\textrm{max}}$ for a given number of atoms (and hence minimises computational cost).

In the results section below we discuss in more detail the various tests we performed to determine the best way of modelling solvent occupancy. The key set of RMC configurations  contained 18\,720 atoms, of which 12\,960 are part of the framework and 5\,760 are part of the included solvent.  We note that the configuration contained six different atom types (C, D, H, N, O, Zn), giving rise to a total of 21 partial $g_{ij}(r)$ contributions---we understand this to be the largest number refined within {\sc rmcprofile} to date. A number of soft constraints and restraints were applied throughout the refinement process: closest-approach constraints stopped atom pairs from being separated by unphysically short distances; `distance-window' constraints maintained the framework connectivity without prejudicing the bond-bending and bond-stretching terms [Table~\ref{table1}];\cite{Goodwin:2005} and, finally, a set of empirical geometric restraints were applied to maintain the geometry of the isonicotinate moiety.\cite{Bennett:2010fk}

\begin{table}
\caption{\label{table1} `Distance window' parameters \emph{d}$_{min}$, \emph{d}$_{max}$ used in our Zn(ISN)$_2$ RMC refinements and the corresponding mean pair separations {\emph{\=d}} (\AA) and their standard deviations $\sigma$.}
\begin{center}
\begin{tabular}{@{\extracolsep{5mm}}lcccc}      % Alignment for each cell: l=left, c=center, r=right
\hline\hline Atom pair&$d_{\textrm{min}}$ (\AA)&$d_{\textrm{max}}$ (\AA)&$\bar d$ (\AA) & $\sigma$ (\AA)\\\hline
O--H	  & 0.8 & 1.1	& 1.006		& 0.076\\
C--D   & 0.9 & 1.2	& 1.039		& 0.072\\
C--C     & 1.1 & 1.7    &  1.431       & 0.062\\
C--N    & 1.1 & 1.7         & 1.321	& 0.055 \\
C--O 	  & 1.1 & 1.7 		& 1.268	& 0.040 \\
Zn--O  & 1.8  & 2.4		& 2.004		& 0.128 \\
Zn--N  & 1.8  & 2.4		& 2.081		& 0.104 \\\hline\hline
\end{tabular}
\end{center}
\end{table}

The refinement process included fitting the experimental $F(Q)$ ($0.6\leq Q\leq40$\,\AA$^{-1}$), $I(t)$ and $G(r)$ ($1.6\leq r\leq25.2$\,\AA) functions. We note here that the minimum value of $r$ used for the $G(r)$ fitting is larger than the nearest-neighbour bond length for the vast majority of solvents, including H$_2$O. A total of five independent RMC refinements were performed in parallel; each refinement was allowed to continue until no further improvements in the fits to data were observed. The absolute atomic coordinates differ amongst the five final configurations, but the corresponding fits to data are essentially identical. Wherever possible, our results are averaged over all five configurations.

\section{Results}
\subsection{Average structure}\label{averagestructure}

\begin{figure}
\includegraphics{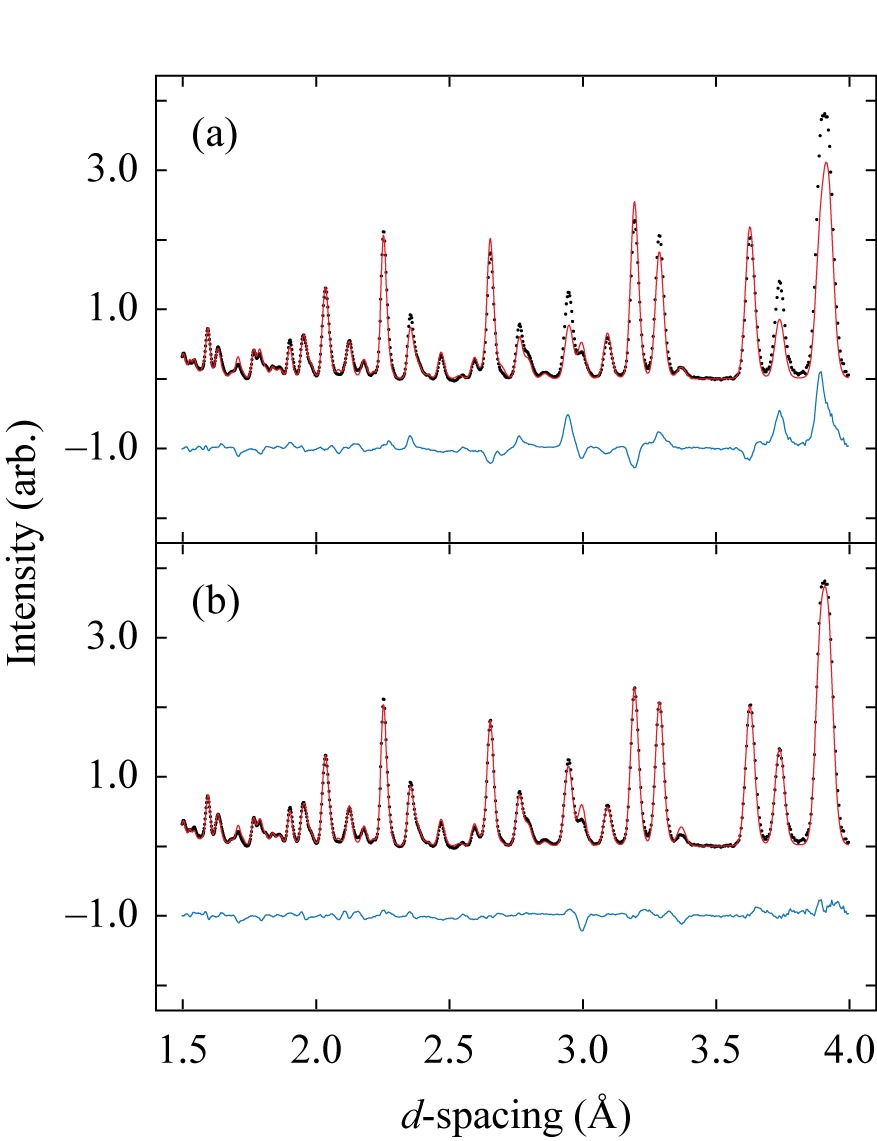}
\caption{Representative Reitveld fits to the 10\,K neutron diffraction pattern of Zn(ISN)$_2$: (a) using an evacuated framework model, and (b) using a model containing four water molecules per formula unit.  Experimental data are given as small filled circles, the fitted profile is shown as a solid red line, and the difference (fit$-$data) is shown beneath each curve.\label{fig2}}
\end{figure}

We used as the basis of our Rietveld refinements the models of Refs.~\onlinecite{Wu:2009} and \onlinecite{Sun:2002}, both of which include a number of water molecules within the framework pore structure. Our first set of refinements involved simply removing the solvent component and refining framework coordinates. We obtained essentially identical fits for both $P3_1$ and $P6_2$ models, neither of which was entirely convincing [Fig.~\ref{fig2}(a)]. A considerable improvement to the quality of these fits was obtained by including scattering density within the framework pores. We tested a number of different ways of modelling this scattering density---including the coordinates of Refs.~\onlinecite{Wu:2009} and \onlinecite{Sun:2002}---but found the best fits were obtained for a slightly different structure model containing four H$_2$O molecules per formula unit [Fig.~\ref{fig2}(b)]. The corresponding crystallographic details are summarised in Table~\ref{table2}.

It is certainly feasible that the sample absorbed atmospheric H$_2$O during setup of the GEM experiment or that some H$_2$O or DMSO (used in the synthesis) remained from incomplete desolvation during sample preparation. Because all of the various models for adsorbed solvent we tested involved large displacement parameters and fractional occupancies of multiple equivalent sites, we are certainly not claiming that the diffraction intensities are sufficiently sensitive to identify the chemical composition of the included component. Instead, we were keen to assess the extent to which the structural parameters of the framework itself were affected by the different models. The positions and atomic displacement parameters obtained for the `four H$_2$O' and solvent-free models are given in Table~\ref{table2}; their close correspondence illustrates that the framework geometry is robustly defined by the diffraction data, even if those of the solvent component are not. In all our refinements, we found no evidence to support symmetry lowering to the $P3_1$ model of Ref.~\onlinecite{Sun:2002}.

\begin{table}
\begin{center}
\caption{\label{table2}Crystallographic parameters, atomic coordinates 
and isotropic equivalent displacement parameters determined using 
Rietveld refinement of neutron scattering data for Zn(ISN)$_2$. The ISN 
coordinates are given as the centre-of-mass of the ligand. The first set 
of positional and displacement parameters correspond to the final 
structural model containing four H$_2$O molecules per formula unit; the 
second (marked with an asterisk) correspond to the solvent-free model 
discussed in the text.}
\begin{tabular}{lcccc}
\hline\hline
\multicolumn{1}{l}{Crystal system}&\multicolumn{4}{l}{Hexagonal}\\
\multicolumn{1}{l}{Space group}&\multicolumn{4}{l}{$P6_2$}\\
\multicolumn{1}{l}{$a$ (\AA)}&\multicolumn{4}{l}{15.53615(25)}\\
\multicolumn{1}{l}{$c$ (\AA)}&\multicolumn{4}{l}{6.13556(26)}\\
\multicolumn{1}{l}{$V$ (\AA$^3$)}&\multicolumn{4}{l}{1282.54(6)}\\
\multicolumn{1}{l}{$Z$}&\multicolumn{4}{l}{3}\\
\multicolumn{1}{l}{$T$ (K)}&\multicolumn{4}{l}{10}\\
\hline\hline
Atom&$x$&$y$&$z$&$U_{\rm iso}$ (\AA$^2$)\\\hline
Zn&0.5& 0 &0.1319(16) & 0.0199(22)\\
ISN&0.26272(8)&0.79764(9)&0.4963(12)&0.0229(3)\\
H$_2$O1&0.7702(11)&0.8083(6)&0.310(4)&0.438(5)\\
H$_2$O2&0.8143(8)&1.0246(7)&0.3744(20)&0.438(5)\\\hline
Zn$^\ast$&0.5& 0 &0.132(3)&0.016(4)\\
ISN$^\ast$&0.26305(15)&0.79501(17)&0.514(3)&0.0204(7)\\\hline\hline
\end{tabular}
\end{center}
\end{table}

\begin{figure}
\includegraphics{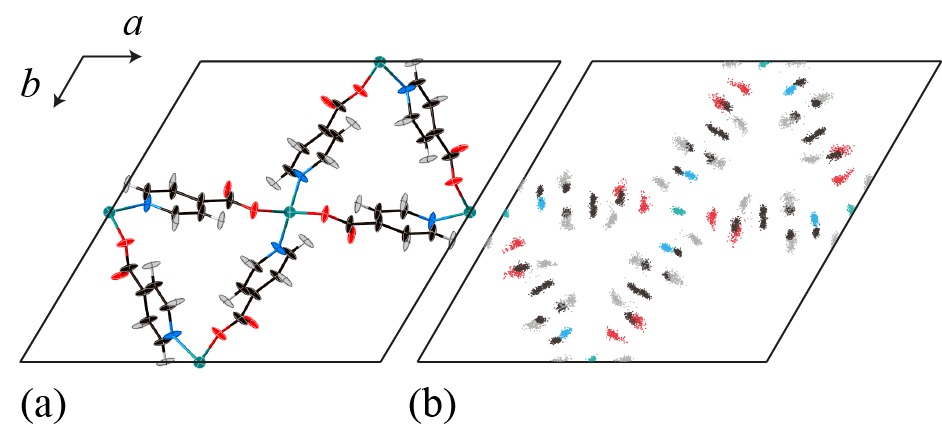}
\caption{Unit cell projections of Zn(ISN)$_2$ viewed down the \emph{c} axis, showing (a) the average structure and (b) the atomic distributions obtained by collapsing the RMC configuration onto a single unit cell.  In both cases, water molecules have been removed for clarity.\label{fig3}}
\end{figure}

It was possible to refine reliably a set of anisotropic displacement parameters for the ISN ligand. The values obtained correspond to thermal ellipsoids that are strongly elongated in a direction perpendicular to the plane of the ligand itself [Fig.~\ref{fig3}(a)]. This sort of behaviour is not uncommon for molecular framework materials, since the lowest-energy vibrational modes usually involve transverse displacements of the bridging ligand; examples include the cyanide bridges of Zn(CN)$_2$ (Ref.~\onlinecite{Goodwin_2005}) and the terephthalate linkers in MOF-5 (Ref.~\onlinecite{Lock_2010}). What is unusual here is the magnitude of the thermal ellipsoids given that the data were collected at 10\,K. We note that by using a wide range of GEM detector banks our refinements include data at sufficiently low $d$-spacing values to give confidence in the determination of anisotropic displacement parameters. The issue of the magnitude of displacement parameters is one to which we return in Section~\ref{discussion}.

\subsection{Local structure}

\begin{figure}
\includegraphics{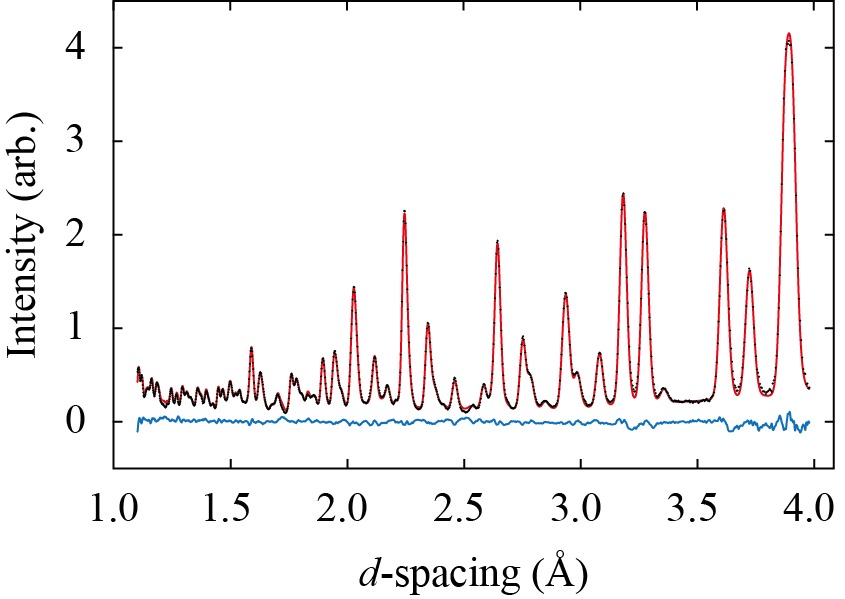}
\caption{RMC fit to the Bragg profile with the experimental data points given as points, and the RMC fits (solid lines) obtained using RMCPROFILE as described in the text.  The difference curve is shown beneath the fit.\label{fig4}}
\end{figure}

With the strong preference in Rietveld refinements for a structure model containing four water molecules per unit cell, our RMC study of local structure in Zn(ISN)$_2$ used this same model as its starting point. RMC refinement gave fits to the time-of-flight Bragg intensity function $I(t)$ [Fig.~\ref{fig4}] that were essentially identical to those obtained by Rietveld refinement. As anticipated, the solvent molecules were strongly disordered in our final RMC configurations and consequently our analysis focusses on the local structure of the Zn(ISN)$_2$ framework itself.

By `collapsing' the atomic coordinates of each refined RMC configuration onto a single unit cell, it is possible to visualise the average structure model to which these configurations correspond [Fig.~\ref{fig3}(b)]. There is a clear similarity to the results of our Rietveld refinement. In particular, there are large anisotropic displacements of the ISN ligand that reflect well the displacement parameters determined above. Whereas the Rietveld refinement employed rigid-body constraints and a single set of displacement parameters for all atoms within the ISN ligand, the RMC refinement allows the scattering distribution for each atomic site to assume whatever form is demanded by the data. We find that the differences amongst distributions for the various atoms in the ISN ligand are small, and certainly in all cases the displacements are much stronger in a direction perpendicular to the ligand plane.

\begin{figure}
\includegraphics{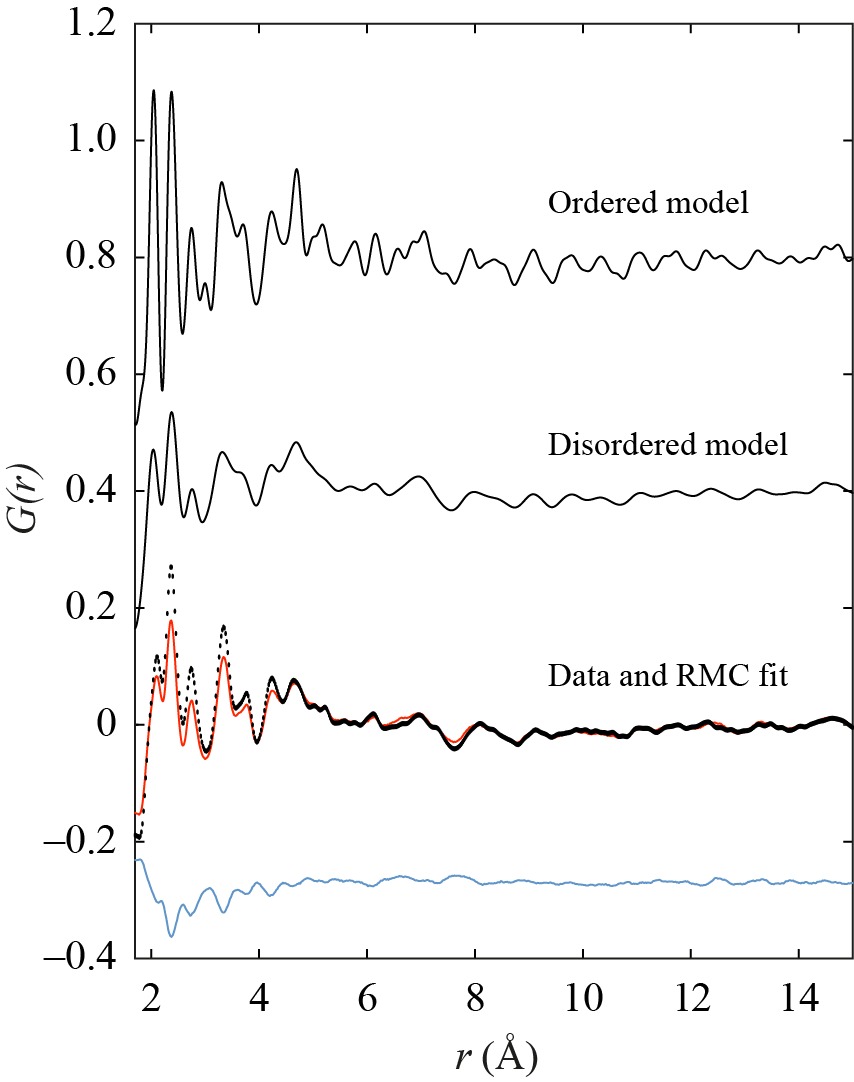}
\caption{Calculated and experimental $G(r)$ functions for Zn(ISN)$_2$. (Bottom) Experimental $G(r)$ data (filled black circles) and RMC fit (red line). (Middle) Calculated $G(r)$ function for the static disordered model obtained from Rietveld refinement. (Top) Calculated $G(r)$ function for the low-energy-dynamics model described in the text.\label{fig5}}
\end{figure}

Given the ambiguity regarding included solvent composition, it is not at all surprising that we found the lowest-$r$ region of the $G(r)$ particularly difficult to normalise robustly. For most of the $G(r)$, the contribution from a disordered solvent network would be expected to be a smoothly-varying function of $r$; however, at the lowest values of $r$ there must exist well-defined contributions corresponding to the interatomic separations within individual solvent molecules. Consequently our RMC fits to $G(r)$ were performed only for $r\geq1.7$\,\AA\ in order that the particular method employed for modelling solvent inclusion (in this case four H$_2$O molecules per unit cell) would not affect refinement of the framework. The fits obtained are not at the quantitative level usually achieved using {\sc rmcprofile}, but nonetheless all qualitative features of the $G(r)$ are certainly captured well [Fig.~\ref{fig5}].

We were struck by the absence of pronounced features in the PDF beyond $r\simeq5$\,\AA, and especially so for data collected at a temperature of 10\,K. Even for a system with a disordered component, the framework itself would be expected to give rise to a well-structured $G(r)$ function: the superposition of two functions---one strongly varying and the other weakly varying---still varies strongly. Consequently, the $G(r)$ function could be considered consistent with a large degree of disorder in the framework geometry of Zn(ISN)$_2$.

\begin{figure}
\includegraphics{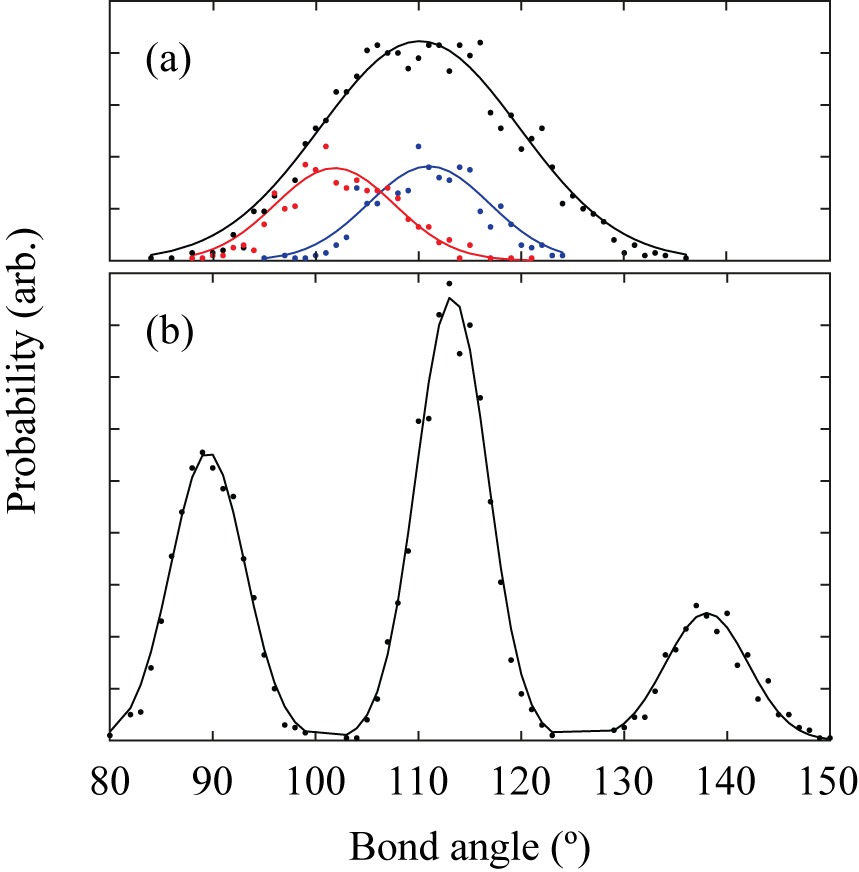}
\caption{ RMC bond angle distributions: (a) intratetrahedral angles within the [ZnN$_2$O$_2$] coordination environment (N--Zn--N in blue, O--Zn--O in red, and N--Zn--O in black); (b) intratetrahedral X--Zn--X angles reflecting geometric flexing of the framework itself.\label{fig6}}
\end{figure}

It is straightforward to determine bond-angle distributions directly from the RMC configurations, and we concentrate here on two particular distribution functions. The first corresponds to the angles within the [ZnN$_2$O$_2$] coordination tetrahedra [Fig.~\ref{fig6}(a)]. The second type of angle is based on the centres of mass of the ISN ligands (which we represent by the symbol `X'): then the X--Zn--X angle distribution [Fig.~\ref{fig6}(b)] is analogous to the O--Si--O tetrahedral angle distribution in $\alpha$-quartz itself. We find in both cases that the distributions are very broad, suggesting that Zn(ISN)$_2$ is flexible both on the scale of the framework itself and also in terms of the individual coordination polyhedra.

The degree of structural distortion can be quantified further using the language of geometric algebra (GA).\cite{Wells_2002} Analysis of RMC configurations using the GA-based code {\sc gasp}\cite{Wells_2002,Wells_2002b,Wells_2004} allows the atomic displacements to be understood in terms of (i) translations, (ii) rotations, and (iii) distortions of fundamental geometric units. In this particular system, there are two types of geometric unit of special interest. The first corresponds to the [ZnN$_2$O$_2$] coordination tetrahedra, the behaviour of which reflects the rigidity of Zn--N and Zn--O bonding interactions. The second type of unit involves the centres of mass of the ISN ligands: here the behaviour of the [ZnX$_4$] tetrahedra describes the flexing and deformation of the framework structure as a whole.

\begin{figure}
\includegraphics{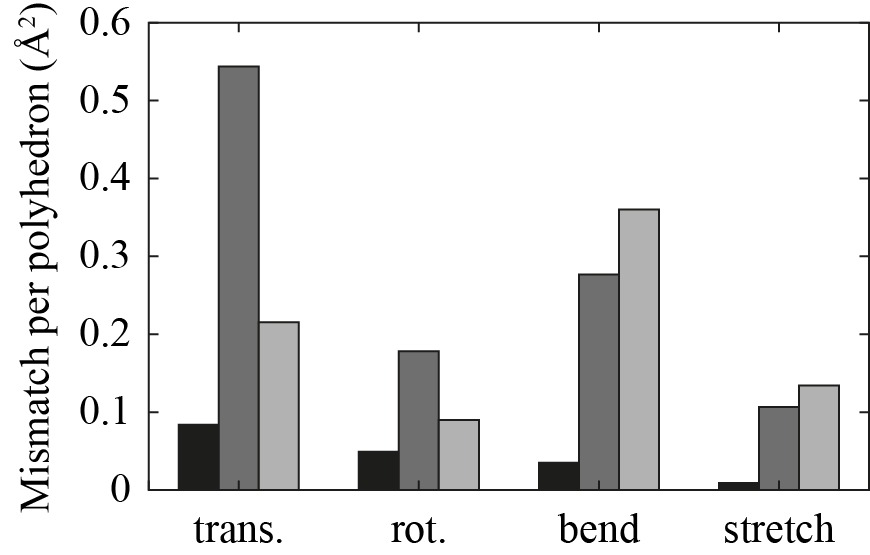}
\caption{Average mismatch scores obtained using {\sc gasp} for $\alpha$-quartz (black), the [ZnO$_2$N$_2$] coordination environment in Zn(ISN)$_2$ (dark grey), and the [ZnX$_4$] framework geometry of Zn(ISN)$_2$ (light grey).  The bar chart shows the type of distortions present in the polyhedra: (left--right) translations, rotations, bond angle bending and bond stretching.\label{fig7}}
\end{figure}

In Fig.~\ref{fig7} we plot the relative components of translations, rotations, and deformation (stretching and bending) determined for the two types of polyhedral unit across our five RMC configurations. What we find is that the translational component is very large indeed for the Zn coordination polyhedra. These translations are evidently allowed by flexing of the Zn--ISN--Zn linkages, reflected in the large angle bending component of the [ZnX$_4$] values. These distortions will correspond to increased displacement of the Zn atoms away from their average sites, and also an apparent increase in the transverse component of the anisotropic displacement parameters for the ISN ligands. These two effects are precisely those observed in Rietveld refinement.

Finally, we put these results in context by comparing the magnitude of the translation, rotation and distortion components with those obtained for an RMC refinement of $\alpha$-quartz itself (also at 10\,K) [Fig.~\ref{fig7}].\cite{Tucker:2001_aquartz} Not only are the rigid-body-type translations and rotations many times larger for Zn(ISN)$_2$, but the distortions are also an order of magnitude more extreme. This result is consistent with the general perception that MOF-type materials are substantially more flexible than their oxide-based counterparts.

\section{Discussion}\label{discussion}

That the ISN ligands in Zn(ISN)$_2$ are displaced in a transverse direction to a very large degree even at 10\,K seems clear both from the average- and local-structure analysis performed so far. The central question is whether or not the magnitude of this displacement is consistent with low-energy vibrational motion of the linkages or with static disorder. Here we make use of the following lattice dynamical formalism which links magnitude of displacement with phonon mode energies:
\begin{equation}
\langle u_j^{2}\rangle=\frac{\hbar}{m_j\omega_{\textrm E}}\left[\frac{1}{2} + n(\omega_{\textrm E},T)\right],\label{u2}
\end{equation}
where $n(\omega_{\textrm E},T)=1/\{\exp[\hbar \omega_{\textrm E}/k_{\textrm B}T] - 1\}$ is the Bose-Einstein occupation number and $m$ the atomic mass.

Because 10\,K is low with respect to typical phonon frequencies, the displacements in equation Eq.~\eqref{u2} will be dominated at this temperature by the energy of the first dispersionless branch of the phonon spectrum. Phonon frequencies have not been determined for Zn(ISN)$_2$, but we might expect that the relevant span of energies will be roughly similar to that in other MOFs; \emph{e.g.}\ MOF-5, the structure of which is also based on Zn$^{2+}$ centres but connected via terephthalate ligands (similar in size to isonicotinate).\cite{Zhou:2006} The first dispersionless branch in MOF-5 is calculated to occur at $\omega=4.6$\,THz. Using this value as a conservative estimate for $\omega_{\textrm E}$ and substituting  into Eq.~\eqref{u2} one obtains $\langle u_{\textrm{Zn}}^2\rangle=0.0113$\,\AA$^2$ and $\langle u_{\textrm{ISN}}^2\rangle=0.0059$\,\AA$^2$. Comparing these values to the $U_{\textrm{iso}}$ values in Table~\ref{table2}, it is clear that the experimental Zn and ISN displacements are, respectively, two and four times larger than can be accounted for by thermal motion alone. This is clear evidence that the displacements observed for Zn(ISN)$_2$ are consistent only with static disorder of the framework.

As a further check of this conclusion, we calculated using {\sc pdfgui} (Ref.~\onlinecite{Farrow_2007}) the $G(r)$ functions expected from the Rietveld model of Section~\ref{averagestructure} using both the as-refined $U_{ij}$/$U_{\textrm{iso}}$ parameters (\emph{i.e.}\ with static disorder) and using equivalent parameters re-scaled according to the $\langle u_i^2\rangle$ values estimated above (\emph{i.e.}\ for a comparable model where displacements are due only to low-energy vibrational modes). The results of both calculations are shown in Fig.~\ref{fig5}, from which it is clear that (i) the dynamic model gives a PDF that is much more typical of 10\,K data, and (ii) the static disorder model reflects well the actual experimental $G(r)$.

In conclusion, we have used a combination of Rietveld and RMC refinement of neutron total scattering data to study the local and average structure of the metal--organic framework zinc(II) isonicotinate. While our study has unquestionably been complicated by the inclusion of hydrogen-containing solvent molecules within the pore structure of the framework, we do find robust evidence for the existence of large-scale static disorder in the material at 10\,K. The role of structural flexibility in the driving unusual dynamic properties of MOFs is increasingly appreciated; what we find here is that the same flexibility may also affect the ground-state structural properties of the same materials.

\section*{Acknowledgements}

This research was supported financially by the EPSRC (grant EP/G004528/2) and the ERC (project 279705), and by the STFC in the form of access to the GEM instrument at ISIS.


\begin{references}

\bibitem{Hoskins_1990}
B. F. Hoskins and R. Robson, J. Am. Chem. Soc. {\bf 112}, 1546 (1990).

\bibitem{Lehnert_1980}
R. Lehnert and F. Seel, {\it Z. Anorg. Allg. Chem.} {\bf 464}, 187 (1980).

\bibitem{Tian_2003}
Y. Q. Tian, C. X. Cai, X. M. Ren, C. Y. Duan, Y. Xu, S. Gao, and X. Z. You, {\it Chem. Eur. J.} {\bf 9}, 5673 (2003).

\bibitem{Banerjee_2008}
R. Banerjee, A. Phan, B. Wang, C. Knobler, H. Furukawa, M. O'Keeffe, and O. M. Yaghi, {\it Science} {\bf 319}, 939 (2008).

\bibitem{Cheetham:2007kx}
A. K. Cheetham and C. N. R. Rao, Science {\bf 318}, 58 (2007).


\bibitem{Ma:2010vn}
S. Ma and H.-C. Zhou, Chem. Commun. {\bf 46}, 44 (2010).

\bibitem{Dybtsev:2006ys}
D. N. Dybtsev, A. L. Nuzhdin, H. Chun, K. P. Bryliakov, E. P. Talsi, V. P. Fedin and K. Kim,  Angew. Chem. Int. Ed. {\bf 45}, 916 (2006).

\bibitem{Halder:2002zr}
G. J. Halder, C. J. Kepert, B. Moubaraki, K. S. Murray and J. D. Cashion, Science {\bf 298},1762 (2002).

\bibitem{Qiu:2009ly}
S. Qiu and G. Zhu, Coord. Chem. Rev. {\bf 253}, 2891 (2009).

\bibitem{Li_1999}
H. Li, M. Eddouadi, M. O'Keeffe and O. M. Yaghi, Nature {\bf 402}, 276 (1999).

\bibitem{Zhou:2008uq}
W. Zhou, H. Wu, T. Yildirim, J. R. Simpson and A. R. Hight Walker, Phys. Rev. B {\bf 78}, 054114 (2008).

\bibitem{Wu:2008kx}
Y.  Wu, A. Kobayashi, G. J. Halder, V. K. Peterson, K. W.
  Chapman, N. Lock, P. D. Southon and C. J. Kepert, Angew. Chem. Int. Ed. {\bf 47}, 8929 (2008).
  
 \bibitem{Bennett:2010fk}
T. D. Bennett, A. L. Goodwin, M. T. Dove, D. A. Keen, M. G.
  Tucker, E. R. Barney, A. K. Soper, E. G. Bithell, J.-C. Tan and
  A. K. Cheetham, Phys. Rev. Lett. {\bf 104}, 115503 (2010).

\bibitem{Serre:2002vn}
C. Serre, F. Millange, C. Thouvenot, M. Nogues, G. Marsolier, D. Louer and G. Ferey, J. Am. Chem. Soc. {\bf 124}, 13519 (2002).

\bibitem{Mary:1996}
T. A. Mary, J. S. O. Evans, T. Vogt and A. W. Sleight, Science {\bf 272}, 90 (1996).

\bibitem{Wragg:2008}
D. S. Wragg, R. E. Morris and A. W. Burton, Chem. Meter. {\bf 20}, 1561 (2008).

\bibitem{Rodriguez:2009}
E. E. Rodriguez, A. Llobet, T. Proffen, B. C. Melot, R. Seshadri, P. B. Littlewood and A. K. Cheetham, J. Appl. Phys. {\bf 105}, 114901 (2009).

\bibitem{Lister:2004}
S. E. Lister, I. R. Evans, J. A. K. Howard, A. Coelho and J. S. O. Evans. Chem. Commun. {\bf 36}, 2540 (2004) .

\bibitem{Barea_2003}
E. Barea, J. A. R. Navarro, J. M. Salas, N. Masciocchi, S. Galli and A. Sironi, Polyhedron {\bf 22}, 3051 (2003).

\bibitem{Amirjalayer:2008fk}
S. Amirjalayer and R. Schmid, J. Phys. Chem. C {\bf 112}, 14980 (2008).

\bibitem{Jhon:2007uq}
Y. H. Jhon, M. Cho, H. R. Jeon, I. Park, R. Chang, J. L. C. Rowsell and J. Kim, J. Phys. Chem. C {\bf 111}, 16618 (2007).

\bibitem{Wu:2009}
Y. Wu, D. Li, F. Fu, L. Tang, J. Wang and X.-G. Yang, J. Coord. Chem. {\bf 62}, 2665 (2009).

\bibitem{Sun:2002}
J. Sun, L. Weng, Y. Zhou, J. Chen, Z. Chen, Z. Liu and D. Zhao, Angew. Chem. Int. Ed. {\bf 41}, 4471 (2002).

\bibitem{Tucker:2007}
M. G. Tucker, D. A. Keen, M. T. Dove, A. L. Goodwin and Q.  Hui, J. Phys.: Condens. Matt. {\bf 19}, 335218 (2007).

\bibitem{Williams:1997}
W. G. Williams, R. M. Ibberson, P. Day and J. E. Enderby, Phys. B: Condens. Matt. {\bf 241-243}, 234 (1997).

\bibitem{Day:2004}
P. Day, J. Enderby, W. Williams, L. Chapon, A. Hannon, P. Radaelli and
  A. Soper, Neutron News {\bf 15}, 19 (2004).

\bibitem{AlexC:2005}
A. C. Hannon, Nucl. Instr. Meth. Phys. Res. A {\bf 551}, 88 (2005).

\bibitem{Keen:2001fk}
D.A. Keen, J. Appl. Crystallogr. {\bf 34}, 172 (2001).

\bibitem{Farrow:2007}
C. L. Farrow, P. Juhas, J. W. Liu, D. Bryndin, E. S. Bozin, J. Bloch, T. H. Proffen and S. J. L. Billinge, J. Phys.: Condens. Matt. {\bf 19}, 335219 (2007).


\bibitem{VonDreele:2000}
A. C. Larson and R. B. Von Dreele, {\it General Structure Analysis System (GSAS)} (Los Alamos National Laboratory report, 2000).

\bibitem{Toby:2001}
B. H. Toby, J. Appl. Crystallogr. {\bf 34}, 210 (2001).

\bibitem{Tucker:2001}
M. G. Tucker, M. P. Squires, M. T. Dove and D. A. Keen, J. Phys.: Condens. Matt. {\bf 13}, 403 (2001).

\bibitem{Goodwin:2005} %check
A. L. Goodwin, M. G. Tucker, E. R. Cope, M. T. Dove and D. A. Keen, Phys. Rev. B {\bf 72}, 214304 (2005).


\bibitem{Dove:2002kx}
M. T.  Dove, M. G. Tucker and D. A. Keen, Eur. J. Mineral. {\bf 14}, 331 (2002).



\bibitem{Zwijnenburg:2006fk}
M. A. Zwijnenburg, R. Huenerbein, R. G. Bell and F. Cora, J. Solid State Chem. {\bf 179}, 3429 (2006).

\bibitem{Goodwin_2005}
A. L. Goodwin and C. J. Kepert, Phys. Rev. B {\bf 71}, 140301 (2005).

\bibitem{Lock_2010}
N. Lock, Y. Wu, M. Christensen, S. L. Cameron, V. K. Peterson, A. J. Bridgeman, C. J. Kepert and B. B. Iversen, J. Phys. Chem. C {\bf 114}, 16181 (2010).

\bibitem{Wells_2002}
S. A. Wells, M. T. Dove and M. G. Tucker, J. Phys.: Condens. Matt. {\bf 14}, 4567 (2002).

\bibitem{Wells_2002b}
S. A. Wells, M. T. Dove, M. G. Tucker and K. Trachenko, J. Phys.: Condens. Matt. {\bf 14}, 4645 (2002).

\bibitem{Wells_2004}
S. A. Wells, M. T. Dove and M. G. Tucker, J. Appl. Crystallogr. {\bf 37}, 536 (2004).

\bibitem{Tucker:2001_aquartz}
M. G. Tucker, D. A. Keen, and M. T. Dove, Mineral. Mag. {\bf 65}, 489 (2001).

\bibitem{Zhou:2006}
W. Zhou and T. Yildirim, Phys. Rev. B {\bf 74}, 180301 (2006).
  
\bibitem{Farrow_2007}
C. L. Farrow, P. Juhas, J. W. Liu, D. Bryndin, E. S. Bozin, J. Bloch, T. Proffen and S. J. L. Billinge, J. Phys.: Condens. Matt. {\bf 19}, 335219 (2007).


\end{references}
\end{document}